\documentclass[prb,twocolumn,showpacs,preprintnumbers,amsmath,amssymb]{revtex4}
\usepackage{graphicx}% Include figure files
\usepackage{dcolumn}% Align table columns on decimal point
\usepackage{bm}% bold math

\begin{document}

\title{Collective Josephson vortex dynamics in a finite number of intrinsic Josephson
junctions}

% Force line breaks with \\

\author{Myung-Ho Bae$^1$}
\author{Jae-Hyun Choi$^1$}
\author{Hu-Jong Lee$^{1,2,*}$}

\affiliation{$^1$Department of Physics, Pohang University of
Science
and Technology, Pohang 790-784, Republic of Korea}%
\affiliation{$^2$National Center for Nanomaterials Technology,
Pohang 790-784, Republic of Korea}

\date{\today}

\begin{abstract}
We report the experimental confirmation of the collective
transverse plasma modes excited by the Josephson vortex lattice in
stacks of intrinsic Josephson junctions in
Bi$_{2}$Sr$_{2}$CaCu$_{2}$O$_{8+x}$ single crystals. The
excitation was confirmed by analyzing the temperature ($T$) and
magnetic field ($H$) dependencies of the multiple sub-branches in
the Josephson-vortex-flow region of the current-voltage
characteristics of the system. In the near-static Josephson vortex
state for a low tunneling bias current, pronounced
magnetoresistance oscillations were observed, which represented a
triangular-lattice vortex configuration along the $c$ axis. In the
dynamic vortex state in a sufficiently high magnetic field and for
a high bias current, splitting of a single Josephson vortex-flow
branch into multiple sub-branches was observed. Detailed
examination of the sub-branches for varying $H$ field reveals that
sub-branches represent the different modes of the Josephson-vortex
lattice along the $c$ axis, with varied configuration from a
triangular to a rectangular lattices. These multiple sub-branches
merge to a single curve at a characteristic temperature, above
which no dynamical structural transitions of the Josephson vortex
lattice is expected.

\end{abstract}

\pacs{74.72.Hs, 74.50.+r, 74.78.Fk, 85.25.Cp }
%\keywords{Suggested keywords}%Use showkeys class option if keyword
                              %display desired
\maketitle

\section{Introduction}

It has been known that stacked intrinsic Josephson junctions
(IJJs) form in naturally grown Bi$_{2}$Sr$_{2}$CaCu$_{2}$O$_{8+x}$
(Bi-2212) single crystals.\cite{Kleiner} The collective Josephson
plasma oscillation, manifested by the electromagnetic resonant
absorption in such Josephson-coupled layered superconductors, has
provided the key to understanding the superconducting state and
the vortex phases\cite{long} forming in the materials. Two kinds
of collective Josephson plasma modes, \emph{longitudinal} and
\emph{transverse}, exist in such a system of an \emph{infinite
number} of staked IJJs.\cite{CJP} The longitudinal plasma mode
oscillation, excited in an externally applied $c$-axis-oscillating
electric field, propagates along the $c$ axis of a stack of IJJs
with the inter-layer phase difference of a junction being unform
in a planar direction. The resonant absorption of externally
applied microwaves by the excitation of the longitudinal modes
enables one to directly obtain the Josephson plasma frequency,
$\omega_p$. This, in turn, provides information on the interlayer
Josephson coupling strength and the $c$-axis superfluid density in
Bi-2212.\cite{Gaifullin} By contrast, the transverse plasma modes
can be excited by the oscillating magnetic fields in the $ab$
plane. The resulting induced current along the $c$ axis generates
the oscillating electric field that resonates with the transverse
Josephson plasma oscillation.

On the other hand, a system with a \emph{finite number} ($N$) of
stacked IJJs, described by linearized coupled sine-Gordon
differential equations with the inductive inter-junction coupling,
exhibits \emph{collective transverse plasma} (CTP) eigen modes,
with the eigen frequencies lying between those of the longitudinal
and the transverse modes found for a system of an infinite number
of stacked junctions.\cite{Inductive,Machida1} The corresponding
dispersion relation is expressed as
\begin{equation}
\omega_{n}(k)=\omega_p\sqrt{1+\frac{\lambda_c^2k^2}
{1+(2\lambda_{ab}^2/sD)(1-\mbox{cos}(\frac{\pi n}{N+1}))}}.
\end{equation}
Here, $\lambda_{ab}$ $(\lambda_c)$ is the $c$-axis ($ab$-plane)
London penetration depth, $N$ the number of the junctions in a
stack, $s$=0.3 nm ($D$=1.2 nm) the thickness of a superconducting
(insulating) layer, $k$(=$2\pi m/L$; $L$=the length of a stack;
$m$=the number of the vortices present in a junction) the wave
number of the plasma oscillation.\cite{Inductive} The mode index
$n$ runs from 1 to $N$.

The CTP modes can be excited by the moving Josephson vortex
lattice forming in high magnetic fields, which are alternative
solutions of the coupled sine-Gordon equations.\cite{Machida} The
temporal oscillation of the inter-layer phase difference due to
the moving Josephson vortex lattice, with the frequency matching
with that of any CTP modes, excites a resonant plasma oscillation.
The Josephson vortex lattice also transforms its spatial
configuration along the $c$ axis in accordance with the $c$-axis
standing-wave modes of the plasma oscillation. The resonance of
Josephson vortex lattice to the CTP modes is revealed in the form
of the multiple Josephson vortex-flow branches in the tunneling
current-voltage ({\it I-V}) characteristics of a stack of
IJJs.\cite{MJV,Bae1}

Experimental observation of the multiple CTP modes in a stack of a
finite number of IJJs has often been attempted in a mesa or an
S-shaped stack of IJJs fabricated on the surface of a large stack
of IJJs, which is called a pedestal or a basal
layer.\cite{Hech,Hirata} Since each superconducting bilayer in a
stack of IJJs is much thinner than the $c$-axis London penetration
depth $\lambda_{ab}$, Josephson vortices in a usual mesa or an
S-shaped-stack structure are in a strong inductive coupling to
those in the basal layer that contains a large number of IJJs by
itself, although measurements are intended to be made on a finite
number of stacks in the structure of interest. Thus, the stack of
IJJs with a coupling to the basal layers corresponds to a system
containing nearly an infinite number of junctions, which does not
support the CTP mode formation along the $c$ axis. This situation
significantly affects the dynamics of the Josephson vortices in
the mesa or the S-shaped stack itself. This may explain the usual
failure of observing the multiple resonating modes in those
structures.

In this paper, we present the static and dynamic states of
Josephson vortices in stacks of IJJs, each sandwiched between two
Au electrodes without the basal layer, thus consisting of a
genuinely finite number of IJJs. In the near-static state in a low
tunneling bias current on a single Josephson vortex-flow branch,
pronounced magnetoresistance oscillations were observed with
$h/4e$ magnetic flux period. This half-flux-quantum periodic
resistance oscillations indicate that the Josephson vortex lattice
in the near-static state constitutes a triangular lattice along
the $c$ axis. In the dynamic state in a high tunneling bias
current, on the other hand, a single Josephson vortex-flow branch
splits into multiple sub-branches. Detailed examination of the
sub-branches for varying magnetic fields and temperatures revealed
that the observed sub-branches were caused by the resonance
between the Josephson vortex lattices and the CTP modes.

\begin{figure}[t]
\begin{center}
\leavevmode
%h=here, t=top, b=bottom, p=separate figure page
\includegraphics[width=0.6\linewidth]{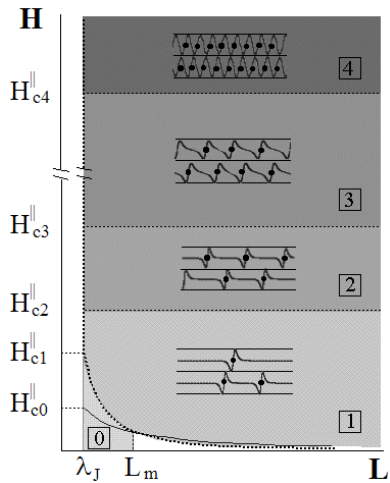}
\caption{The phase diagram of the Josephson vortex configuration
for different ranges of transverse magnetic field and the length
of the intrinsic-Josephson-junction stack. A schematic figure in
the each region represents the supercurrent distributions in
layered structures, where the centers of the Josephson vortices
are indicated by black dots.}
\end{center}
\end{figure}

\section{Basic concepts}

The Josephson vortex configuration in stacked IJJs of Bi-2212
single crystals is determined by the magnetic flux density in the
junctions and the junction size.\cite{Koshelev} Josephson vortices
are introduced to the insulating layer of a junction when its
length $L$ is longer than the Josephson penetration depth,
$\lambda_{J}$ $[$=$
\sqrt{\Phi_{0}/2\pi\mu_{0}j_{c}(D+2\lambda_{ab}^{2}/s)}]$, where
$\Phi_0($=$h/2e)$ is a flux quantum. One can identify various
characteristic magnetic fields in relation with the dynamics of
Josephson vortices [see Fig. 1]. For a stack of junctions of
length $L$ Josephson vortices start entering the junctions above a
characteristic field,\cite{Clem, Fistul} $H_{c1}^{||}$
[=$(\Phi_0\lambda_c/L^2\lambda_{ab})$ln$(\lambda_{ab}/D)]$. This
field can be higher than the nominal value of one flux quantum
threading a junction of length $L$, that is $H_{c0}^{||}$
(=$\Phi_0/(s+D)L$), especially for a junction comparable to
$\lambda_J$. Even above $H_{c0}^{||}$ the tunneling screening
current prevents an external magnetic field from entering the
junctions until it reaches $H_{c1}^{||}$. Thus, for a junction
with its length in the range of $\lambda_J<L<L_m$
[$\equiv$$\lambda_J$ln$(\lambda_{ab}/(s+D)$], one can observe the
usual oscillatory Fraunhofer modulation of the tunneling critical
current as a function of an external magnetic field of
$H<H_{c1}^{||}$, although the junction size is larger than
$\lambda_J$ (Region 0).\cite{Latyshev1}

The Josephson vortices entering the junctions for $H>H_{c1}^{||}$
form a regular lattice as the magnetic field reaches the
value\cite{Ikeda} $H_{c2}^{||}$ (=$1.4\Phi_{0}/2\pi (s+D)
\lambda_{J})$. In this field range (Region 1), since the
field-induced vortices exist in arbitrary configurations in a
given magnetic field, disordered multiple vortex-flow subbranches
are generated in the {\it I-V} characteristics. In this region,
the number of vortices in a junction and the propagation
velocities of different vortex configurations are not correlated
with each other.\cite{Krasnov} Thus, in this field range, the
vortex-flow subbranches in the {\it I-V} characteristics are not
related to vortex mode velocities\cite{Sakai,Machida1} or
geometrical resonances.\cite{Krasnov2}

In an external field above $H_{c2}^{||}$ (Region 2), Josephson
vortices start forming a triangular lattice. Recent observation of
the oscillatory vortex-flow resistance arising from the
interaction between the vortex lattice and the crystal boundary
potential clearly confirmed the existence of the triangular vortex
lattice\cite{Ooi, Machida2} above $H_{c2}^{||}$.

On the other hand, the resonance between the Josephson vortex
lattice and CTP modes is well known in this vortex dynamics
system.\cite{Machida1} The collective resonance revealed as
voltage peaks (in the case of the voltage bias) or voltage jumps
(in the case of the current bias) has been searched extensively
both theoretically and
experimentally.\cite{Inductive,Hech,Koshelev2} The collective
resonance has been observed in the dense-vortex state
corresponding to the field range of $H>H_{c3}^{||}$
[=$\Phi_0/2\lambda_J(s+D) $], where the shortest inter-vortex
spacing becomes comparable to the diameter of a Josephson vortex
$2\lambda_{J}$ (Region 3). Thus, Josephson vortices in Regions 2
and 3, although they form triangular lattices in both regions in a
static state without a tunneling bias current (Fig. 1), behave
differently in the vortex-flow state in the presence of a finite
tunneling bias current. In the dynamic state, Josephson vortices
in Region 2 remains in a triangular lattice, while those in Region
3 transform their lattice configuration to fit the plasma
propagation modes in the stacked junctions.\cite{Kleiner3} The
non-Josephson-like emission from the Josephson-vortex
motion\cite{Hech2} and the Shapiro resonance steps in the
Josephson vortex-flow state\cite{Latyshev2} have been observed in
this dense-vortex state. These results confirm the existence of
the coherent motion of the Josephson vortex lattice through the
entire thickness of stacked (intrinsic) Josephson junctions.

For the highest characteristic field $H_{c4}^{||}$
[=$\Phi_0/\lambda_J(s+D) $ ($\sim$4 T for Bi-2212)], the shortest
inter-vortex spacing becomes comparable to the Josephson
penetration depth,\cite{Bulaevskii} $\lambda_{J}$. In this region
(Region 4), the Josephson current along the length of a junction
distributes almost sinusoidally so that the collective resonance
behavior between the vortex motion and the plasma modes is
expected to be stronger than in a field lower than $H_{c4}^{||}$.

\section{Experiment}

\begin{figure}[b]
\begin{center}
\leavevmode
%h=here, t=top, b=bottom, p=separate figure page
\includegraphics[width=1\linewidth]{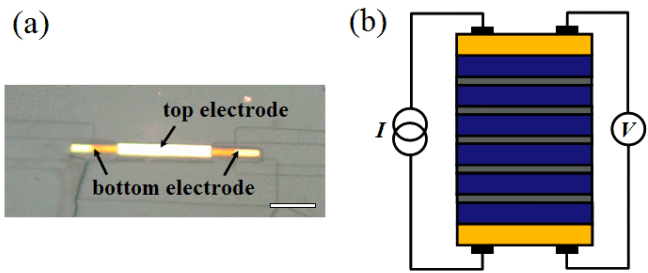}
\caption{(color online) (a) An optical micrograph of a sample
during the fabrication process. The scale bar represents the
length of 10 $\mu$m. (b) Schematics of sample and measurement
configurations.}
\end{center}
\end{figure}

As-grown slightly overdoped Bi-2212 single crystals were prepared
by the conventional self-flux method. We fabricated, using the
double-side cleaving technique,\cite{Wang1} three samples of IJJs
sandwiched between two Au electrodes at the top and the bottom of
a stack of IJJs without the basal part [Fig. 2(b)]. This
structure, which is in contrast to the usual mesa or the
S-shaped-stack structure fabricated on the surface of a single
crystal with a large basal layer, is more coincident with the
model systems usually adopted in the theoretical
analysis.\cite{Sakai} The geometry with the basal layer eliminated
by the double-side cleaving enables one to investigate the
Josephson vortex dynamics in coupled IJJs without the interference
from the vortex motion in the basal layer. Furthermore, the top
and the bottom Au-film leads in a stack help the formation of the
standing plasma oscillation modes along the $c$ axis, while
forming a transmission guide of the plasma oscillations along the
length of the stack.

The samples containing sandwiched stacks were prepared in the
following way.\cite{Bae2} A single crystal was first glued on a
sapphire substrate using negative photoresist and was cleaved
until an optically clean and flat surface was obtained. Then, a
100-nm-thick Au film was thermally deposited on top of the crystal
to protect the surface of the crystal and to make a good ohmic
contact. A mesa structure was then prepared by photolithographic
patterning and Ar-ion-beam etching. The surface of the patterned
mesa was fixed to another sapphire substrate using negative
photoresist and the basal layer of the mesa was subsequently
cleaved away. This process, called the double-side-cleaving
technique, was developed by Wang {\it et al}.\cite{Wang1} A
100-nm-thick Au film was again deposited on this freshly cleaved
crystal surface, leaving a stack of IJJs sandwiched between two Au
electrodes. A few-$\mu$m-long portions on both sides of the
sandwiched stack were etched away subsequently [see Fig. 2(a)] to
get the bottom Au electrode exposed for $c$-axis transport
measurements as schematically depicted in Fig. 2(b). Finally, a
300-nm-thick Au-extension pad was attached on each of the Au
electrode. The lateral sizes of three samples were 13.5 $\times$
1.4 $\mu$m$^2$ [SP1], 15 $\times$ 1.4 $\mu$m$^2$ [SP2] and 16
$\times$ 1.5 $\mu$m$^2$ [SP3].

The magnetic field was aligned in parallel with the plane of
junctions within the resolution of 0.01 degree to minimize the
pinning of Josephson vortices by the pancake vortices that can
form in CuO$_2$ bilayers due to the vertical component of a
misaligned field. The alignment was done, in a field of 4 T and at
a temperature around 60 K, by tuning to the angle that gives the
maximum Josephson-vortex-flow resistance or the minimum pining of
the Josephson vortices by the pancake vortices generated by the
angle misalignment.\cite{Hech} To minimize the external noise,
transport measurements were performed with a low-pass filter
connected to each electrode located at room temperature.

\begin{table}[b]
\caption{\label{tab:tableI}Sample parameters. $N$ is the total
number of intrinsic Josephson junctions in a stack, $\lambda_{J}$
the Josephson penetration depth,
$H_{c3}^{||}$=$\Phi_0/2\lambda_J(s+D)$ ($s$ and $D$ the thickness
of the superconducting and the insulating layers, respectively),
$c_m$ the maximum propagation velocity of vortices in Region 3,
and $c_1$ the velocity of the fastest collective plasma mode
estimated by the RCSJ model.}
\begin{ruledtabular}
\begin{tabular}{ccccccc}
sample&stack size &$N$&$\lambda_J$&
$H_{c3}^{||}$&$c_m$&$c_1$\\
 number& ($\mu$m$^2$)& & ($\mu$m)& (T)&($\times$10$^5$ m/s)&($\times$10$^5$ m/s)  \\
\hline
SP1 &13.5$\times$1.4 &35 &0.25 &2.8 &3.8&3.5\\
SP2 &15$\times$1.4 &22 &0.32 &2.1 &3.9&3.7\\
SP3 &16$\times$1.5 &45 &0.24 &2.9 &4.2&3.5\\
\end{tabular}
\end{ruledtabular}
\end{table}

\section{Results and discussion}

\subsection{Static and dynamic state}

\begin{figure}[t]
\begin{center}
\leavevmode
%h=here, t=top, b=bottom, p=separate figure page
\includegraphics[width=0.8\linewidth]{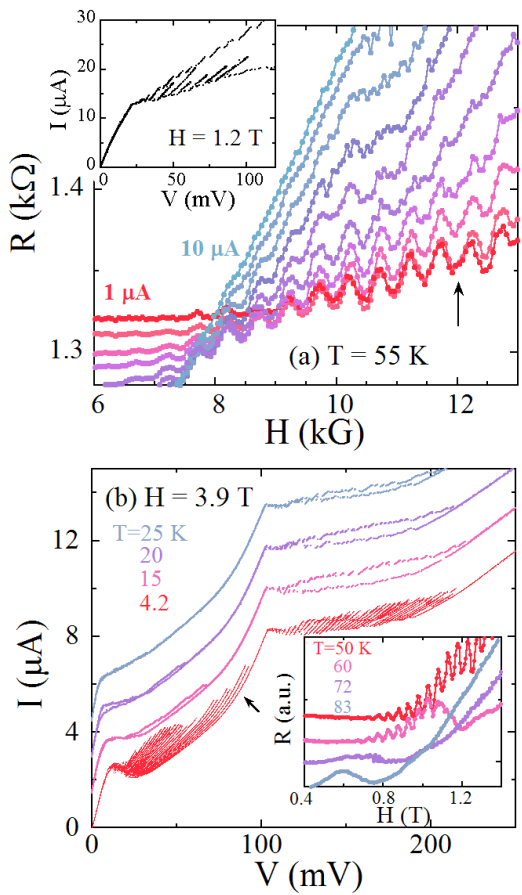}
\caption{(color online) (a) The Josephson vortex-flow resistance,
$R$ as a function of an external magnetic field $H$ for SP1 at
$T$=55 K in bias currents from 1 $\mu$A to 10 $\mu$A at intervals
of 1 $\mu$A. (b) Temperature dependence of the
Josephson-vortex-flow sub-branches and quasiparticle branches of
SP1 at $H$=3.9 T. Inset of (a): a single Josephson vortex-flow
branch and quasiparticle tunneling branches in $H$=1.2 T at $T$=55
K. Inset of (b): the temperature dependence of $R$($H$) in a bias
current of 3 $\mu$A. Curves in the main panel and the inset of (b)
are shifted vertically for clarity.}
\end{center}
\end{figure}

Fig. 3(a) shows the oscillation of Josephson-vortex-flow
resistance of SP1 [$N$=35] as a function of external magnetic
fields at $T$=55 K in bias currents from 1 $\mu$A to 10 $\mu$A at
intervals of 1 $\mu$A. The field period of the oscillation,
$H_p$=510 G is one half of $H_0$ (=$\Phi_0/L(s+D)$) through the
junction area of length $L$=13.5 $\mu$m and thickness $(s+D)$=1.5
nm. The oscillation of the Josephson-vortex-flow resistance is
known to be caused by the interaction between the triangular
Josephson vortex lattice along the $c$ axis and the vortex-flow
boundary potential at the junction edges.\cite{Ooi,KS,Machida2}
The observation of the Josephson-vortex-flow resistance
oscillation in our sandwiched stack without the basal layers
indicates that, as the S-shaped-stack geometry, our sandwiched
geometry also supports the Josephson-vortex triangular lattice in
the steady state [see the inset of Fig. 4(a)]. The oscillation in
Fig. 3(a) is reduced by increasing bias current as the bias
current tilts the boundary potential.\cite{Machida3} The
oscillatory-magnetoresistance behavior is also suppressed
completely at $T$ $\sim$80 K near $T_c$ (=92 K) as in the inset of
Fig. 3(b), which is related to melting of the near-static
Josephson vortex lattice.\cite{Hu}

The inset of Fig. 3(a) shows the single Josephson-vortex-flow
branch (the low-bias region) and quasiparticle branches (the
high-bias region) in $H$=1.2 T at $T$=55 K. The field, belonging
to Region 2 in Fig. 1, corresponds to the position indicated by
the arrow in the main panel. In this field region, the
single-curve Josephson-vortex-flow branch caused by the triangular
lattice persists even down to 4.2 K (not shown). In Fig. 3(b),
{\it I-V} characteristics in a field of Region 3 at $T$=25 K also
show a single Josephson vortex-flow branch corresponding to the
triangular lattice. Table I shows the characteristic field value
of $H_{c3}^{||}$ of each sample.

Contrary to the field range of Region 2, however, the single
Josephson-vortex-flow branch in Region 3 evolves into the multiple
Josephson-vortex-flow sub-branches with decreasing temperature.
Fig. 3(b) shows the temperature dependence of the
Josephson-vortex-flow branches, where the multiple sub-branches
start appearing at $T$$\sim$20 K and become clearer with
decreasing temperature. At $T$=4.2 K, a single branch representing
the triangular lattice below the current bias of $I_b$$\sim$2
$\mu$A splits into multiple Josephson-vortex-flow sub-branches for
the higher current bias. The Josephson vortex flow, then, transits
to the McCumber quasiparticle tunneling for $I_b$ above 8 $\mu$A.
The multiple Josephson-vortex-flow sub-branches rapidly suppress
with increasing temperature above 4.2 K and disappear completely
as the temperature reaches 25 K. On the contrary, the
quasiparticle-tunneling branches are robust to the temperature
variation and persist up to $T$=25 K. This contrasting temperature
dependencies of two different regions strongly imply that the
multiple branches in the two regions are of different physical
origins.

\begin{figure}[t]
\begin{center}
\leavevmode
%h=here, t=top, b=bottom, p=separate figure page
\includegraphics[width=0.82\linewidth]{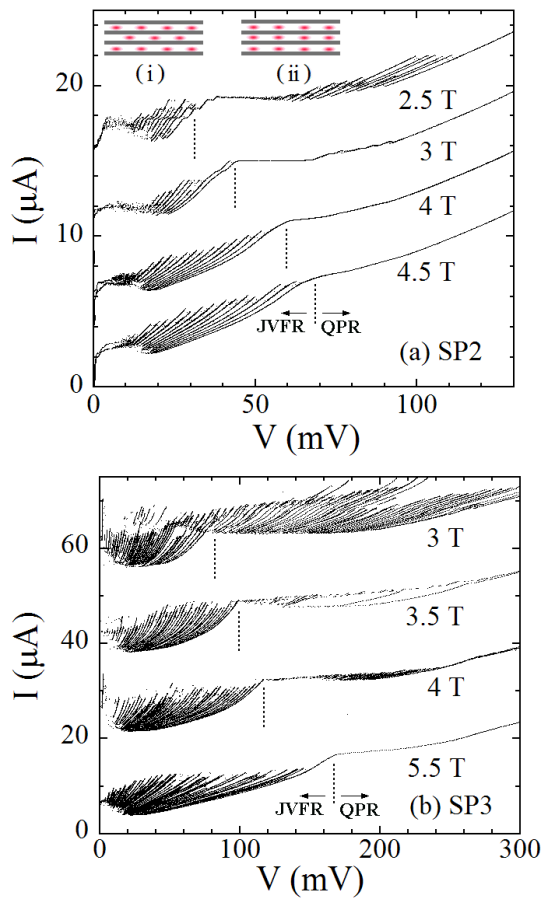}
\caption{(color online) The magnetic field dependencies of the
Josephson vortex-flow sub-branches and the McCumber branches of
(a) SP2 and (b) SP3 at $T$=4.2 K. The {\it I-V} characteristics in
all the figures are shifted vertically for clarity. A voltage
$V_m$ denoted by a vertical dotted line in each set of curves
demarcates the Josephson vortex flow region (JVFR) from the
McCumber quasiparticle-tunneling region (QPR). Inset of (a): (i)
the triangular lattice and (ii) the rectangular lattice
configurations of the Josephson vortex lattice along the $c$ axis.
The contact resistance was subtracted numerically in the main
panels of (a) and (b).}
\end{center}
\end{figure}

We also observed the separation of the two characteristic
multiple-branch regions in SP2 and SP3. Figs. 4(a) and 4(b) show
the magnetic field dependencies of {\it I-V} characteristics of
two sandwiched stacks, SP2 and SP3, respectively, at 4.2 K. The
number of zero-field quasiparticle-tunneling branches of SP2 and
SP3 were 22 and 45, respectively (not shown), which coincided with
the total number of IJJs in the given stacks. The differentiation
of the multiple branches by the collective Josephson vortex flow
(the low-bias region) from the ones by the quasiparticle tunneling
(the high-bias region) is based on the fact that the delimiting
current, represented by the current values marked by the dotted
vertical lines, stemmed from the zero-field Josephson critical
current. Nonetheless, a finite resistance is present in the region
below the critical bias point because of the Josephson vortex
flow, although the region itself represents the dissipationless
pair tunneling state.

The observed sub-branches for SP2 and SP3 have common features:
(i) a single Josephson-vortex-flow branch splits into multiple
sub-branches for Regions 3 and 4, (ii) the number of the
quasiparticle sub-branches is $\sim$18 and $\sim$42, respectively,
which is close to the number of the junctions in each sample, and
(iii) the sub-branches in the Josephson vortex-flow region become
clearer and wider with increasing the transverse magnetic field
(in contrast to the case of quasiparticle sub-branches, which keep
shrinking with field instead). These three common properties of
sub-branches are consistent with the prediction of the inductive
coupling theory for the electrodynamics of stacked Josephson
junctions.

Since the speed of vortices, $c_m$, is limited by that of the
maximum electromagnetic wave in a junction, Josephson-vortex-flow
state changes to quasiparticle tunneling state near the maximum
cut-off velocity or equivalently by the limiting voltage $V_m$ in
the outermost sub-branch. Using the relation\cite{Bae1} of
$V_{m}$=$NH(s+D)c_m$, we could get $c_m$ as
$\sim$3.9$\times$10$^5$ m/s and $\sim$4.2$\times$10$^5$ m/s for
the corresponding delimiting bias point $V_m$ in SP2 and SP3,
respectively. These values are similar to those obtained from a
Josephson-vortex-flow branch in mesa structures.\cite{Hech} In the
case of the usual mesa or the S-shaped-stack structure with a
basal layer, only a single Josephson-vortex-flow branch used to be
observed below the critical bias point, regardless of the
temperature and the magnetic field intensity. The velocity $c_m$
corresponding to the delimiting value on this single branch has
been regarded as the slowest mode velocity, $n$=$N$. As
illustrated below, however, in our case without the basal layer
$c_m$, corresponding to $V_m$ in the outermost sub-branch,
represents the propagation velocity of the $n$=1 plasma mode.

In the resistively and capacitively shunted junction (RCSJ)
model\cite{Irie} the junction capacitance is obtained by the
relation, $C_j$=$\beta_c \frac{\Phi_0}{2\pi I_c R_n^2}$, where
$\beta_c$ [=$(4I_c/\pi I_r)^2$] is the McCumber parameter, $R_n$
is the normal-state junction resistance and $I_r$ is the return
current. In SP2, $C_j$ is obtained to be about 47 pF, with
$I_c$(=0.2 mA), $I_r$(=4.8 $\mu$A) and $R_n$(=10 $\Omega$). The
Swihart velocity $c_0$=$\sqrt{sA/2\mu_0 C_j \lambda_{ab}^2}$,
estimated from the junction capacitance, is 3.6 $\times$ 10$^4$
m/s, where $\lambda_{ab}$=200 nm and the junction area of $A$=21
$\mu$m$^2$. The resulting fastest mode velocity $c_1$=3.7 $\times$
10$^5$ m/s is in remarkable agreement with the observed value of
$c_m$ in SP2. Experimental values of $c_m$ in Table I are
consistent with the calculated ones of $c_1$ for all three
samples. This indicates that the outermost sub-branch is the
fastest velocity, $n$=1, mode.

\begin{figure}[t]
\begin{center}
\leavevmode
%h=here, t=top, b=bottom, p=separate figure page
\includegraphics[width=0.8\linewidth]{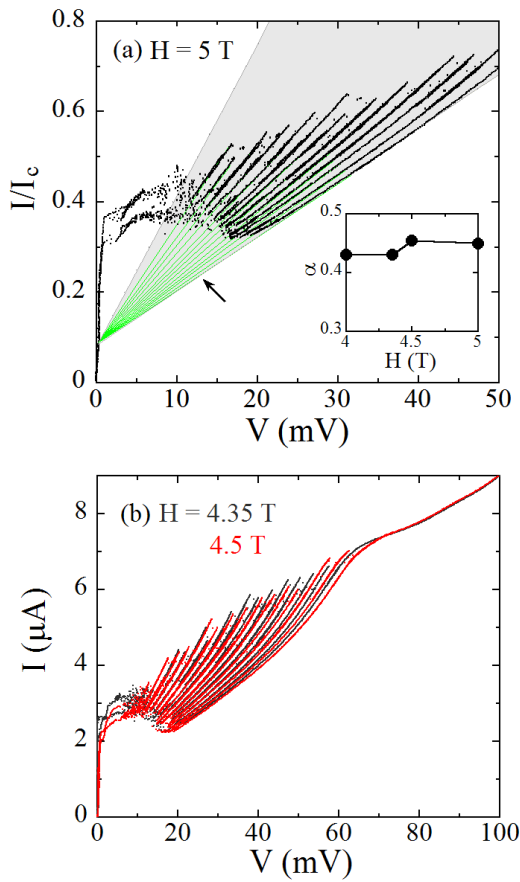}
\caption{(a) Josephson vortex-flow sub-branches for SP2 for $H$=5
T with the current axis normalized by the critical bias current.
Splitting of the low-bias Josephson-vortex-flow curve into
multiple sub-branches indicates the effectiveness of the
inter-junction capacitive coupling. (b) Josephson vortex-flow
sub-branches for SP2 at $H$=4.35 and 4.5 T. The contact resistance
was subtracted numerically in (a) and (b). Inset of (a): magnetic
field dependence of the capacitive coupling parameter, $\alpha$}
\end{center}
\end{figure}

In general, Josephson-vortex resonance branches by the inter-layer
inductive coupling have been predicted to generate current steps
near the resonance voltages on the background of a single
resistive branch.\cite{Inductive} In the present study, however,
Jospehson-vortex resonance to the CTP modes constitutes multiple
sub-branches over almost the entire bias range. A recent theory
for an interlayer inductive coupling hybridized by a capacitive
coupling predicts that the resistive sub-branches corresponding to
the collective resonance modes show linear {\it I-V}
characteristics in the low-bias region with the varied slopes from
one another.\cite{Kim} More spread in the slopes are predicted for
a higher capacitive coupling. Fig. 5(a) shows the
Josephson-vortex-flow sub-branches for SP2 at $H$=5 T. As
predicted by the theory, all the sub-branches in the Josephson
vortex-flow regime exhibit linearities with varied slopes in the
low bias range. The extrapolation of the linear low-bias region
converges to a single point on the current axis, although the
inductive-capacitive hybrid coupling theory predicts they should
converge to the origin. The discrepancy indicates a finite pinning
of Josephson vortex motion due to the presence of any defects in
the stacked junctions or pancake vortices due to any field
misalignment, with the converging current point corresponding to
the depinning current. The inductive-capacitive hybrid coupling
model indicates that the extent of the spread in the low-bias
slope of the resonant branches is proportional to the strength of
the capacitive coupling represented by the parameter $\alpha$ as
$V_n$=$[\frac{I-I_p}{I_c}]\frac{V_c}{1+2\alpha[1-\mbox{cos}(n\pi/N)]}$,
where $n$=1,2,..,$N$. Here, $I_p$ is the depinning current of the
Josephson-vortex, $I_c$ is the tunneling critical Josephson
current at a given applied field. Here $V_c$ is the maximum mode
voltage corresponding to a given bias current as represented by
the curve denoted with an arrow in Fig. 5(a). The gray region
illustrates the best fit to the sub-branches in the low-bias
region. The best-fit value of $\alpha$ turns out to be 0.45 with
$V_c$=4.22 mV. The best-fit values of $\alpha$ were found to be
almost insensitive to magnetic fields as $\alpha$=0.43$\sim$0.45
for $H$=4$\sim$5 T [see the inset of Fig. 5(a)], which is
consistent with the assumption of the capacitive coupling and
reasonably close to the theoretical expectation\cite{Kim} of
0.1$<\alpha<$0.4. Therefore, as shown in the upper inset of Fig.
4(a), we consider that the first ($n$=$N$; the leftmost) and last
($n$=1; the rightmost) sub-branches correspond to (i) the
triangular and (ii) the rectangular Josephson vortex lattices
along the $c$ axis, respectively.

Fig. 5(b) shows the change of Josephson-vortex-flow sub-branches
for SP2 when magnetic field increases from $H$=4.35 T to 4.5 T.
The more changes take place in the voltages between the two
magnetic fields for branches in the higher voltages (branches more
to the right). This behavior can be compared to the dispersion
relation of Eq. (1), which states that the lower-index modes are
more dispersive than the higher-index modes. Thus, the right-most
branches among multiple Josephson-vortex sub-branches in Figs.
3$-$6 are again considered to represent the square vortex lattice
[see the inset of Fig. 4(a)] corresponding to the $n$=1 plasma
excitation mode.

The maximum number of vortex-flow sub-branches counted for various
fields in SP2 is 20, which is two modes smaller than the number of
quasiparticle branches (or, equivalently, than the number of
junctions). Recently, Ryndyk\cite{Ryndyk1} proposed that the
additional dissipation due to the charge-imbalance relaxation due
to quasiparticle injection in a junction can prevent the
structural transformation of Josephson-vortex lattice along the
$c$ axis. In our sample geometry quasiparticles are injected
directly from outside through the outermost two junctions, which
may cause the non-equilibrium charge-imbalance effect in the two
outermost CuO$_2$ bilayer. The resulting suppression of the
configurational transformation of Josephson-vortex lattice reduces
the number of observable resonance modes. For inner CuO$_2$
planes, on the other hand, with the pair tunneling for a bias
below the critical current of the junction, the quasiparticle
injection can be neglected.

\subsection{Temperature dependence}

\begin{figure}[t]
\begin{center}
\leavevmode
%h=here, t=top, b=bottom, p=separate figure page
\includegraphics[width=1\linewidth]{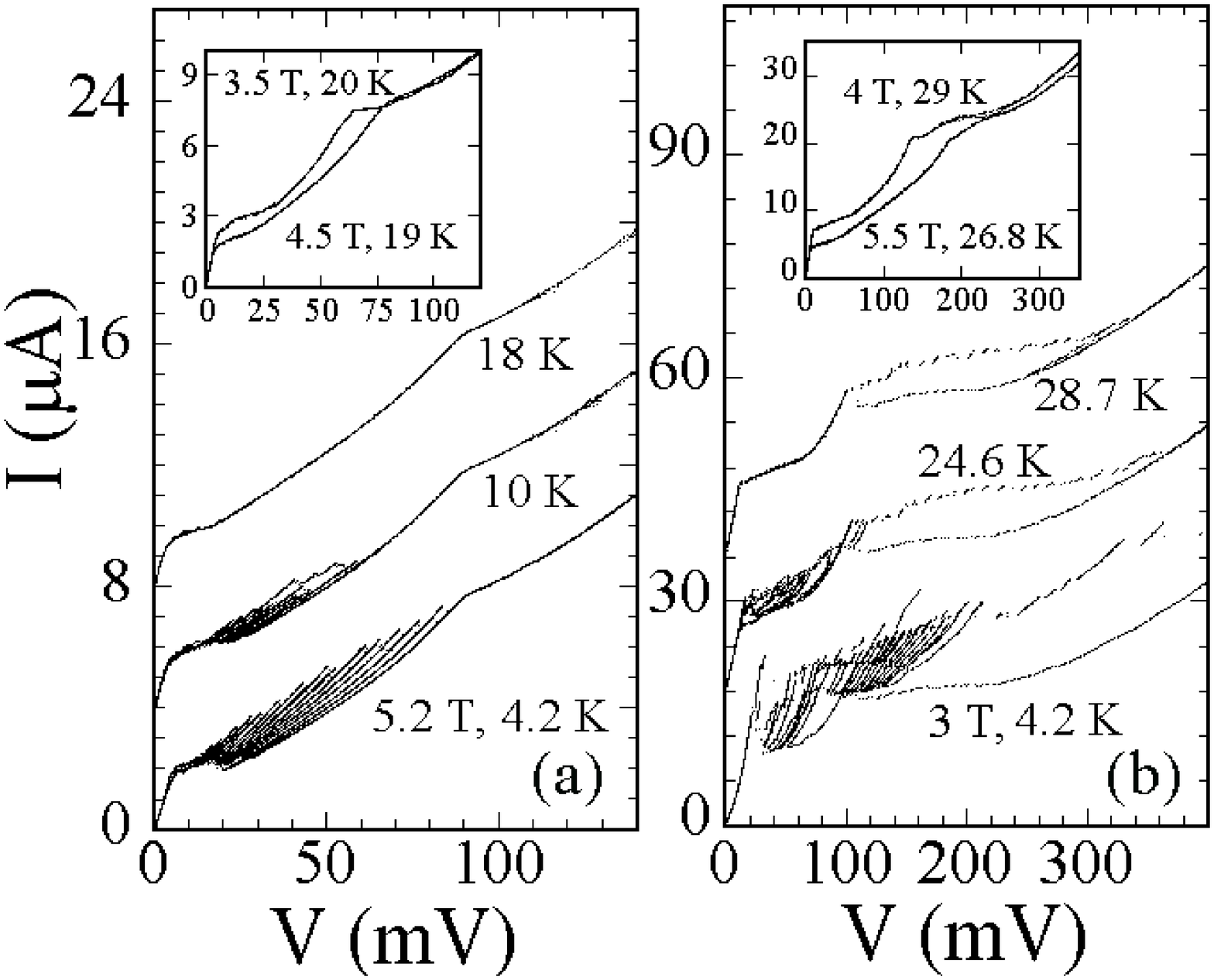}
\caption{Temperature dependence of {\it I-V} characteristics for
(a) SP2 in $H$=5.2 T and (b) for SP3 in $H$=3 T. The curves are
shifted vertically for clarity. The insets of (a) and (b) show a
single Josephson vortex-flow branch above the characteristic
temperature $T_{ch}$ in two different magnetic fields for SP2 and
SP3, respectively.}
\end{center}
\end{figure}

Figs. 6(a) and (b) show the temperature dependencies of {\it I-V}
characteristics for SP2 in $H$=5.2 T and for SP3 in $H$=3 T,
respectively. The temperature dependencies of the sub-branches in
the Josephson vortex-flow region and quasiparticle-tunneling
region are much different from each other. Although not clearly
illustrated in the figures the quasiparticle branches survived up
to temperatures close to $T_c$. The sub-branches in the Josephson
vortex-flow region, however, disappeared at a characteristic
temperature $T_{ch}$, near 18 K and 28 K for SP2 and SP3,
respectively, leaving a single branch as SP1 in Fig. 3(b). These
characteristic temperatures are almost insensitive to the fields
applied [see insets of Figs. 6(a) and (b)].

In the vortex flow state for the magnetic-field range of Regions 3
and 4 the triangular vortex configuration and the multiple-mode
configuration resonating with the collective CTP modes compete
each other. At low-enough temperatures, the triangular vortex
configuration wins in the low-bias region, while the multiple-mode
configuration predominates in the high-bias region. For
$T$$\sim$$T_{ch}$, however, thermal fluctuation and quasiparticle
dissipations prevent the Josephson vortex lattice from resonating
with the collective CTP modes. Then the vortices stay on the
triangular-lattice configuration without a structural
transformation even in the dynamic state.\cite{Ryndyk1} The value
of $V_m$ seems to be insensitive to temperature near $T_{ch}$,
which means that the triangular Josephson vortex lattice can reach
the maximum velocity corresponding to the fastest mode, $n$=1,
without the structural transformation in resonance with the CTP
modes. In this branch above $T_{ch}$, therefore, the lattice
structure may be the modulated triangular lattice but with the
propagation velocity still locked to that of the fastest mode.
\emph{Thus, the single branch above $T_{ch}$ represents the
triangular vortex lattice, while the well-defined outermost
sub-branch below $T_{ch}$ represents the rectangular lattice}.
Accordingly, in Figs. 3(b) and 6, although the rightmost
sub-branch below $T_{ch}$ appears to directly transform into the
single branch above $T_{ch}$, their characteristics correspond to
very different Josephson-vortex configurations.

As mentioned earlier, the single Josephson vortex-flow branch
without sub-branches has been observed in the usual mesa
structure.\cite{Hech} In the usual mesa or the S-shaped-stack
structure the structural transformation from the triangular
lattice to the rectangular lattice in resonance to the CTP modes
is hard to take place because of the strong coupling to the
near-static triangular lattice in the basal layer even at
sufficiently low temperatures. In this case, the velocity of the
Josephson vortex lattice may reach that of the fastest mode
without the structural transformation.\cite{Ryndyk1}

\section{CONCLUSIONS}

The Josephson vortex dynamics was comprehensively studied in
compactly stacked intrinsic Josephson junctions over a wide range
of magnetic fields and temperatures. Measurements were made on
three stacks of intrinsic Josephson junctions, each sandwiched
between two (top and bottom) normal-metallic electrodes, thus
eliminating the interference from the basal layer(s). The {\it
I-V} characteristics for high external magnetic fields (Regions 3
and 4 defined in Fig. 1) showed three states: (i) the near-static
triangular Josephson-vortex state with the formation of a single
branch, (ii) the dynamic vortex state with multiple-mode
sub-branches, and (iii) the McCumber quasiparticle-tunneling state
with quasiparticle sub-branches. In the dynamic vortex state, each
sub-branch corresponded to the different configuration of the
Josephson-vortex lattice along the $c$ axis from a triangular to a
rectangular lattices.\cite{Inductive} With increasing temperature,
the sub-branches disappeared at $T_{ch}$, above which no
configurational transformation of the Josephson-vortex lattice can
be assumed.

Due to the strong inductive coupling to many junctions in the
basal layer(s), the mesa or the S-shaped-stack structure, although
containing a finite number of junctions in the stack under study,
behaves effectively as consisting of an infinite number of
junctions along the $c$ axis. Thus, these structures cannot
support the formation of the multiple Josephson-vortex-flow modes
as predicted theoretically for a finite number of junctions.
Although the $c$-axis boundary condition is not strictly imposed
in these systems the transverse vortex motion is still restricted
by the boundary conditions at the side edges. This is the main
reason why, in the mesa or the S-shaped-stack structure, clear
magnetoresistance oscillations are observed by the interplay
between the near-static triangular Josephson-vortex
configuration\cite{Machida2} and the boundary potential in the
systems, while the transverse Josephson-vortex multiple modes are
not obtained. One should be noted that the multiple collective
Josephson vortex modes can be obtained only in a stack containing
a finite number of (intrinsic) Josephson junctions. The triumph of
this study is that our samples truly satisfied this condition of a
finite number of junctions in a stack and, as a result, both the
near-static Josephson vortex state highly coupled to the edge
potential and the collective transverse Josephson-vortex state
were identified in each junction. This facilitated the detailed
identification of the vortex configurations illustrated in Fig. 1.

\section*{ACKNOWLEDGMENTS}
The authors wish to acknowledge stimulating discussions with R.
Kleiner, H. B. Wang, S. M. Kim, M. Machida, Ju H. Kim, and S.
Sakai. This work was supported by the National Research Laboratory
program administrated by Korea Science and Engineering Foundation
(KOSEF). This paper was also supported by POSTECH Core Research
Program.
\\

*Electronic address: hjlee@postech.ac.kr

\end{document}